\documentclass{aastex}
\usepackage{emulateapj5}
\usepackage{graphicx}
\usepackage{times}

\newcommand{\be}{\begin{equation}}
\newcommand{\en}{\end{equation}}
\def\ltsima{$\; \buildrel < \over \sim \;$}
\def\lsim{\lower.5ex\hbox{\ltsima}}
\def\gtsima{$\; \buildrel > \over \sim \;$}\def\gsim{\lower.5ex\hbox{\gtsima}}

\def\ergs  {\rm \ erg \, s^{-1}}

\def\cmdue {\rm \ cm^{-2}}

\makeatletter

\newenvironment{inlinefigure}{%
\def\@captype{figure}%
\noindent\begin{minipage}{0.999\linewidth}\begin{center}}
{\end{center}\end{minipage}\smallskip}
\makeatother

\slugcomment{Accepted for publication in ApJ Letters}

\begin{document}

\title{The deepest X--ray look at the Universe}

\author{S.~Campana\altaffilmark{1}, A.~Moretti\altaffilmark{1}, 
D.~Lazzati\altaffilmark{2}, G.~Tagliaferri\altaffilmark{1}}

\altaffiltext{1}{Osservatorio Astronomico di Brera, Via E. Bianchi
46, Merate (LC), 23807, Italy.}
\altaffiltext{2}{Institute of Astronomy, University of Cambridge, 
Madingley Road, Cambridge CB3 0HA, UK.}

\authoremail{campana@merate.mi.astro.it}
 
\begin{abstract}
The origin of the X--ray background, in particular at hard (2--10 keV)
energies, has been a debated issue for more than 30 years. The Chandra
deep fields provide the deepest look at the X--ray sky and are the
best dataset to study the X--ray background. We searched the Chandra
Deep Field South for X--ray sources with the aid of a dedicated
wavelet-based algorithm. We are able to reconstruct the Log N--Log S
source distribution in the soft (0.5--2 keV) and hard (2--10 keV)
bands down to limiting fluxes of $2\times 10^{-17}$ erg s$^{-1}$
cm$^{-2}$ and $2\times 10^{-16}$ erg s$^{-1}$ cm$^{-2}$,
respectively. These are a factor $\sim 5$ deeper than previous
investigations. We find that the soft relation continues along the
extrapolation from higher fluxes, almost completely accounting for the
soft X--ray background. On the contrary, the hard distribution shows a
flattening below $\sim 2\times 10^{-14}$ erg s$^{-1}$ cm$^{-2}$.
Nevertheless, we can account for $\gsim 68\%$ of the hard X--ray
background, with the main uncertainty being the sky flux itself.
\end{abstract}


\section{Introduction}

In the soft (1--2 keV) band, ROSAT surveys resolved $70-80\%$ of the
cosmic X--Ray Background (XRB) into discrete sources, at a flux level
of $1 \times 10^{-15}$ erg s$^{-1}$ cm$^{-2}$ (Hasinger et al. 1998).
The great majority of sources brighter than $5 \times 10^{-15}$ erg
s$^{-1}$ cm$^{-2}$ were optically identified with unobscured Active
Galactic Nuclei (AGN; Schmidt et al. 1998; Lehmann et al. 2001). The
origin of the background at higher energies is more
controversial.  The ASCA and {\it Beppo}SAX deep surveys achieved
relatively shallow detection limits of $(3-5)\times 10^{-14}$ erg
s$^{-1}$ cm$^{-2}$, resolving $\sim 25\%$ of the XRB into point
sources (Ogasaka et al. 1998; Cagnoni et al. 1998; della Ceca et
al. 2000; Giommi et al. 2000). Main contributors 
are thought to be absorbed and unabsorbed AGN with a mixture of quasars 
and narrow emission-line galaxies as optical counterparts  
(e.g. Fiore et al. 1999; Akiyama et al. 2000). Recently,
Chandra and XMM-Newton have dedicated long exposures to address this issue 
(Mushotzky et al. 2000;
Giacconi et al. 2001a; Hornschemeier et al.  2000, 2001; Brandt et
al. 2001; Hasinger et al. 2001; Tozzi et al. 2001).  The analysis of
the first 300~ks of the Chandra Deep Field South (CDFS) allowed to reach 
limiting fluxes of $\sim 10^{-16}$ erg s$^{-1}$ cm$^{-2}$ (0.5--2 keV) and
$\sim 10^{-15}$ erg s$^{-1}$ cm$^{-2}$ (2--10 keV; Tozzi et
al. 2001). Similar results were obtained on the Hubble Deep Field
by Chandra observations (Hornschemeier et al. 2000, 2001). These
observations allowed the resolution of $81-95\%$ of the soft (1--2
keV) XRB, confirming the results from the deep ROSAT HRI surveys in
the Lockman hole (Hasinger et al. 1998). In the hard band, the
absolute value of the background is poorly established and, depending
on the adopted normalization, Tozzi et al. (2001) were able to account
for $60-90\%$ of the hard (2--10 keV) XRB.

The increasing angular resolution and source crowding provided by the
current generation of X--ray instruments demands advanced
detection algorithms. One of these is the wavelet transform (WT), a
mathematical tool for multi-scale image analysis. The automatic
detection and characterization of sources in X--ray images by means of
WT-based algorithms were first carried out by Rosati et al.  (1994,
1995). WT-based techniques have been developed and applied also by
other groups to the analysis of ROSAT images (e.g. Slezak et al. 1994;
Grebenev et al.  1995; Damiani et al. 1997; Vikhlinin et al. 1998;
Lazzati et al. 1999). In view of the excellent results so far
obtained, it is natural to extend these techniques to the analysis of
the far richer Chandra images (Freeman et al. 2001; Harnden et
al. 2001: Moretti et al. 2001).  Here we analyzed the full (0.97 Ms)
CDFS (Giacconi et al. 2001b; Rosati et al. 2001) by means of an
advanced detection algorithm originally developed for the analysis of
the ROSAT HRI fields (Lazzati et al. 1999; Campana et al. 1999) based
on the WT (Brera Multiscale Wavelet - BMW). 
At variance with other WT algorithm, our BMW characterizes detected sources 
by means of a decimation scheme (Lazzati et al. 1999). The application and
testing of this algorithm on Chandra fields and on the CDFS in
particular is extensively described in Moretti et al. (2001).

\section{Chandra observations} 

The CDFS consists of eleven observations for a total exposure time of
0.97 Ms.  All exposures were taken with the Chandra X--ray Observatory
(Weisskopf et al. 2000) Advanced CCD Imaging Spectrometer (ACIS-I)
detector (Garmire et al. 2001).  ACIS-I consists of four CCDs arranged
in a $2\times 2$ array for a field of view of $16.9'\times 16.9'$. The
on-axis image quality is $\sim 0.5''$ FWHM, increasing to $\sim 3.0''$
FWHM at $\sim 4.0'$ off-axis.  The data were filtered to include only
the standard event grades 0, 2, 3, 4 and 6 (see Chandra proposers'
observatory guide).  All hot pixels and bad columns were removed. We
removed also flickering pixels with more than two events contiguous in
time (time interval of 3.3 s) and residual cosmic ray events.

The eleven observations were co-added with a pattern recognition
routine to within $0.5''$ r.m.s. pointing accuracy (Moretti et
al. 2001). We restricted our analysis to the ACIS-I area fully exposed
in the eleven observations, selecting a circle with $8'$ radius, for a
total coverage of $\sim 0.056$~deg$^2$.  In order to compare our
results with previous studies we carried out the detection in two
selected energy bands: a 0.5--2 keV soft band and a 2--7 keV hard band
(in the range 7--10 keV the effective area steeply decreases while the
background increases resulting in a lower signal to noise ratio for
X--ray sources; fluxes in the hard 2--7 keV band are extrapolated to
the classical 2--10 keV energy band).  Time intervals during which the
background rate is larger than $3\,\sigma$ over the quiescent level
were removed in each band separately.  This results in a net exposure
time of 942 ks in the soft band and 936 ks in the hard band. The
average background over the fully exposed portion is 0.14 (0.19) counts
per square arcsec in the soft (hard) band.  Following Tozzi et al.  (2001),
the count-rate to flux conversion factors in the 0.5--2 keV and in the
2--10 keV bands were computed using the response matrices at the
aimpoint. A count rate of 1 count s$^{-1}$ in the soft band
corresponds to a flux of $(4.6\pm 0.1) \times 10^{-12}$ erg s$^{-1}$
cm$^{-2}$ and of $(2.9\pm 0.3) \times 10^{-11}$ erg s$^{-1}$ cm$^{-2}$
in the hard band.  These numbers were computed assuming a Galactic
absorbing column of $8 \times 10^{19}$ cm$^{-2}$ and a power law
spectrum with a photon index $\Gamma = 1.4$ (i.e. the slope of the
hard XRB).  The quoted uncertainties are for photon indices in the
range $\Gamma = 1.1 - 1.7$. As shown by Tozzi et al. (2001) a
conversion factor changing with flux, in order to follow the hardening
of the sources, results in a small variation of the Log N--Log S
distribution, at a level of $\sim 5\%$.

A critical parameter in the search of faint sources is represented by
the detection threshold. In the context of detection by wavelet
algorithms this is usually fixed in terms of spurious sources per
field (Lazzati et al. 1998). To compare our results with those
reported by Tozzi et al. (2001), we require to have less than 4.3
spurious sources within the selected area (i.e. 6 spurious sources per
Chandra field).  This threshold has been accurately tested with
simulations and the number of spurious sources is well under control
(Moretti et al. 2001).  This threshold corresponds to a source
significance larger than $4\,\sigma$.

\section{Log N--Log S relations in the soft and hard bands} 

We searched the inner image with the BMW-Chandra detection algorithm
described in Moretti et al. (2001). We detect 239 and 147 sources in
the soft and hard band, respectively. There are 22 sources ($\sim 8\%$
of the total number of detected sources) that are revealed only in the
hard band, and 116 sources ($\sim 44\%$) that are revealed only in the
soft band.

The sky coverage at a given flux, i.e. the area over which a given
source can be detected above our threshold (as due to the varying
vignetting, angular resolution and exposure time), has been derived by
extensive simulations (100 fields per band) in 4 circular annuli (see
Moretti et al. 2001). The inhomogeneities of the ACIS-I field of view,
such as gaps between the CCDs, are dealt with throrough simulations.

The flux distribution of the detected sources suffers from the well
known Eddington bias (e.g. Hasinger et al. 1993). Faint sources can be
detected only if superposed on positive background fluctuations (due
either to a faint contaminating source or a statistical background
fluctuation) and therefore their fluxes are systematically
overestimated by factor as large as 2--3. This bias starts affecting the data 
at a level of $\sim 20$ counts in the soft band ($\sim 10^{-16}\ergs\cmdue$) and 
$\sim 30$ counts in the hard band ($\sim 10^{-15}\ergs\cmdue$). This effect has a 
stronger impact on the 2--10 keV Log N--Log S, due to the higher value 
of the background. Following the approach by Vikhlinin et al. (1995), we
statistically correct for this bias. The completeness of the detection
is then accounted for by the completeness function. The accuracy of
the procedure is assured by extensive simulations (see Moretti et
al. 2001 and Figs. 4--5 therein).

Our simulations show that we are able to recover the number source
distribution down to 4 corrected counts (11 before the bias correction) 
in the inner core of the image, declining to 8 corrected counts
(14 before correction) in the outskirts for the soft band. 
For the hard band the limit on the corrected counts are 5 in the center 
and 10 in the outskirts. 
These counts give a flux limit in the inner region of $2 \times 10^{-17}$ erg s$^{-1}$ 
cm$^{-2}$ in the soft band and $2\times10^{-16}$ erg s$^{-1}$ cm$^{-2}$ 
in the hard band. 

\medskip

\begin{inlinefigure}
\centerline{\includegraphics[width=1.\textwidth]{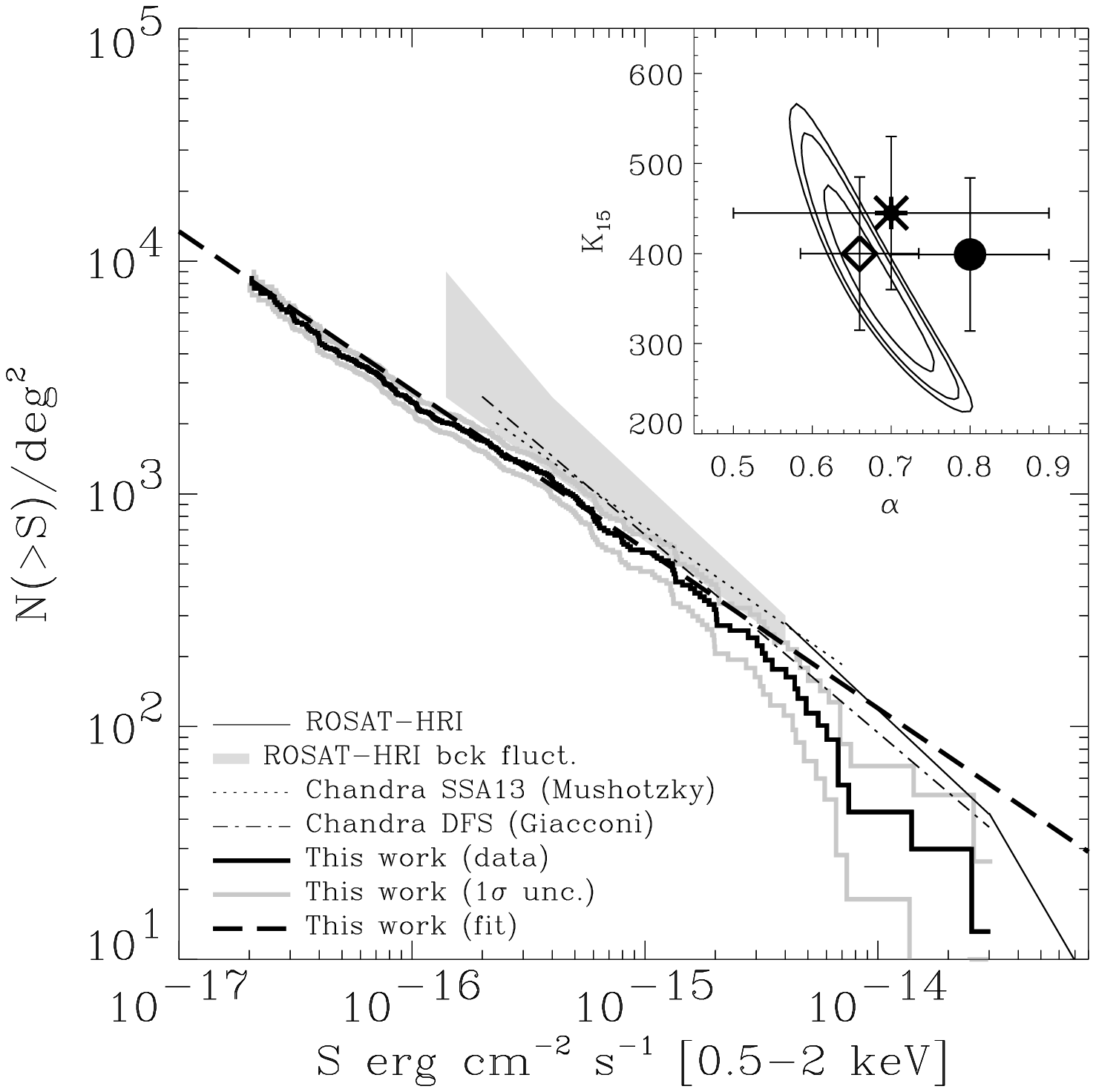}}
\caption{{Soft band Log N--Log S from inner $8'$ radius of the 1~Ms 
observation of the CDFS (thick solid line) in the range
$2\times10^{-17}-3\times10^{-14}$~erg~cm$^{-2}$~s$^{-1}$ for an
average $\Gamma=1.4$ power--law spectrum. The gray thick line shows
the $1\sigma$ confidence region taking into account the number
statistics and the flux conversion uncertainties. The thick dashed
solid line shows the best M--L fit power law, while the other lines
show fits from previous works. The thin solid line shows the
ROSAT--HRI results on the Lockman hole (Hasinger et al. 1998); the
shaded area shows the result of the background fluctuation analysis of
ROSAT--HRI data (Hasinger et al. 1993), while the dotted and
dash--dotted lines show the best fit from the analysis of the SSA13
field Mushotzky et al. (2000) and of the first 120~ks on the CDFS
Giacconi et al. (2001), respectively.  The results on the first 300~ks
of the CDFS (Tozzi et al. 2001) are not shown, since they would have
been virtually indistinguishable from our best fit line. The inset
shows the 1, 2 and $3\,\sigma$ contours from the simultaneous M--L fit
of the slope and normalization of the Log N--Log S. The asterisk,
circle and diamond are the results of Mushotzky et al. (2000),
Giacconi et al. (2001) and Tozzi et al. (2001), respectively.}
\label{fig:lnlss}}
\end{inlinefigure}

\medskip

To characterize the flux distribution of sources detected in the soft
band we performed a maximum likelihood fit to the differential
(unbinned) source flux distribution with a power--law. We obtain as a
best fit:

\begin{equation} 
N(>S) = 360\, \Bigl( {{S}\over{2\times 10^{-15}}}\Bigr) ^{-0.68} \ {\rm cgs}.
\end{equation}

\noindent Fitting the same data, adaptively binning to 10 sources per
bin, we obtain a good fit by means of a $\chi^2$ test (null hypothesis
probability $\sim 10\%$). This assures us of the goodness of the fit.
The error at $90\%$ confidence level for the normalization is
$K_{15}=360^{+41}_{-37}$ ($K_{15}$ means that fluxes are measured in
units of $2\times 10^{-15}$ erg s$^{-1}$ cm$^{-2}$), while the best fit 
value for the slope is $\alpha_{\rm s}=-0.68\pm0.03$ (see also
Fig.~\ref{fig:lnlss}). Both our normalization and slope, are in good
agreement with ROSAT (Hasinger et al. 1998) Chandra (Tozzi et
al. 2001, Rosati et al. 2001) and XMM-Newton (Hasinger et al. 2001) up
to their faintest flux levels.

Due to Galactic contributions and following all previous studies, we
consider the soft XRB in the 1--2 keV band only. Our fitted Log N--Log
S distribution gives an integrated flux of $\sim 5.4 \times 10^{-13}$
erg s$^{-1}$ cm$^{-2}$ deg$^{-2}$ for fluxes fainter than $10^{-15}$
erg s$^{-1}$ cm$^{-2}$. Summing the contribution at higher fluxes
(Hasinger et al. 1998), the total flux of resolved point sources is 
$\sim 3.5 \times 10^{-12} $ erg s$^{-1}$ cm$^{-2}$
deg$^{-2}$.  This accounts for $80\%$ of the unresolved extragalactic
background measured by ROSAT and by a joined ASCA/ROSAT analysis (Chen
et al. 1997).  Extending to even lower fluxes the Log N--Log S
distribution (assuming the same slope) we can make up to $85\%$ of the
soft XRB, at the most.  This is in line with recent predictions for
the soft XRB produced by the warm/hot diffuse intergalactic medium
which may contribute at a $5-15\%$ level (Phillips et al. 2001).  If, instead, 
we assume the lower (by a factor $\sim 0.85$) value measured by ASCA
(Gendreau et al. 1995), then we are able to account for $96\%$ of the flux in the 
soft XRB and fully account for it extending the source distribution at the 
lowest level. 

For the hard band our faint limit can be confidently taken down to
$2\times 10^{-16}$ erg s$^{-1}$ cm$^{-2}$ (see Fig.~\ref{fig:lnlsh}).
In order to be consistent with the results at larger fluxes, we
normalized our distribution to the ASCA data (della Ceca et
al. 2000)\footnote{This normalization of the Log N--Log S distribution
allows to get rid of cosmic variance at bright fluxes. At faint fluxes
this is not true and the small field analyzed can play likely a role,
e.g. in the Chandra observation of the HDF more sources are detected
at the same flux level (Brandt et al. 2001).}.  The Log N--Log S
distribution cannot be described with a single power--law (probability
of the null hypothesis $\sim 0.02\%$, Fig.~\ref{fig:lnlsh}).  We tried
to model the data with a broken power--law and with a
smoothly--jointed broken power--law. The second function gives a much
better agreement with the data and it has the form:

\begin{equation} 
N(>S) = 1.2\times 10^4\, \Bigl[ 
{{(2\times 10^{-15})^{\alpha_{\rm h1}}} \over
{S^{\alpha_{\rm h1}}+S_0^{\alpha_{\rm h1}-\alpha_{\rm h2}}\,
S^{\alpha_{\rm h2}}}} \Bigr] \ {\rm cgs}.
\end{equation}

\noindent 
To match the bright flux end, we fixed the first slope to the ASCA
value, i.e.  $\alpha_{h1}=1.67$ (della Ceca et al. 2000).  The maximum
likelihood fit yields a faint--end slope $\alpha_{\rm h2}=0.58\pm0.03$
and a break flux $S_0=1.7^{+0.3}_{-0.2}\times 10^{-14}$ erg s$^{-1}$
cm$^{-2}$ (all at $90\%$ confidence level for a single parameter; see
Fig.~\ref{fig:lnlsh}).

The total contribution to the XRB is then $1.6\times 10^{-11}$ erg
s$^{-1}$ cm$^{-2}$ deg$^{-2}$. This value has to be compared with the
total (unresolved) XRB of $1.6 \times 10^{-11}$ erg s$^{-1}$ cm$^{-2}$
deg$^{-2}$ from UHURU and HEAO-1 (Marshall et al. 1980), as well as
with the more recent determinations from the {\it Beppo}SAX and ASCA
of $2.4 \times 10^{-11}$ erg s$^{-1}$ cm$^{-2}$ deg$^{-2}$ (Chiappetti
et al. 1998; Vecchi et al.  1999; Ueda et al. 1999; Ishisaki et
al. 2000; Perri \& Giommi 2000), in agreement with the old Wisconsin
measurements (McCammon et al. 1983).  Our results are in good
agreement with the UHURU/HEAO-1 value (considering also the error in
the flux determination due to the uncertainties in the background
slope, which amounts to $10\%$). The slope of the faint end counts
converges slower than logarithmically and thus the faint sources provide
a small contribution to the X--ray background (adding only
$10\%$ more extrapolating down to zero flux).  
In this case we can claim to have fully resolved
the 2--10 keV hard XRB.  On the contrary, if we adopt the {\it
Beppo}SAX and ASCA value, the fraction of resolved sources in the CDFS
lowers to $68\%$. If we extrapolate our best fit distribution to lower
fluxes, we can make up $73\%$ of the entire XRB, at most. If this were the
case a contribution from truly diffuse emission or from a new class of
sources has to be advocated.

\medskip

\begin{inlinefigure}
\centerline{\includegraphics[width=1.0\textwidth]{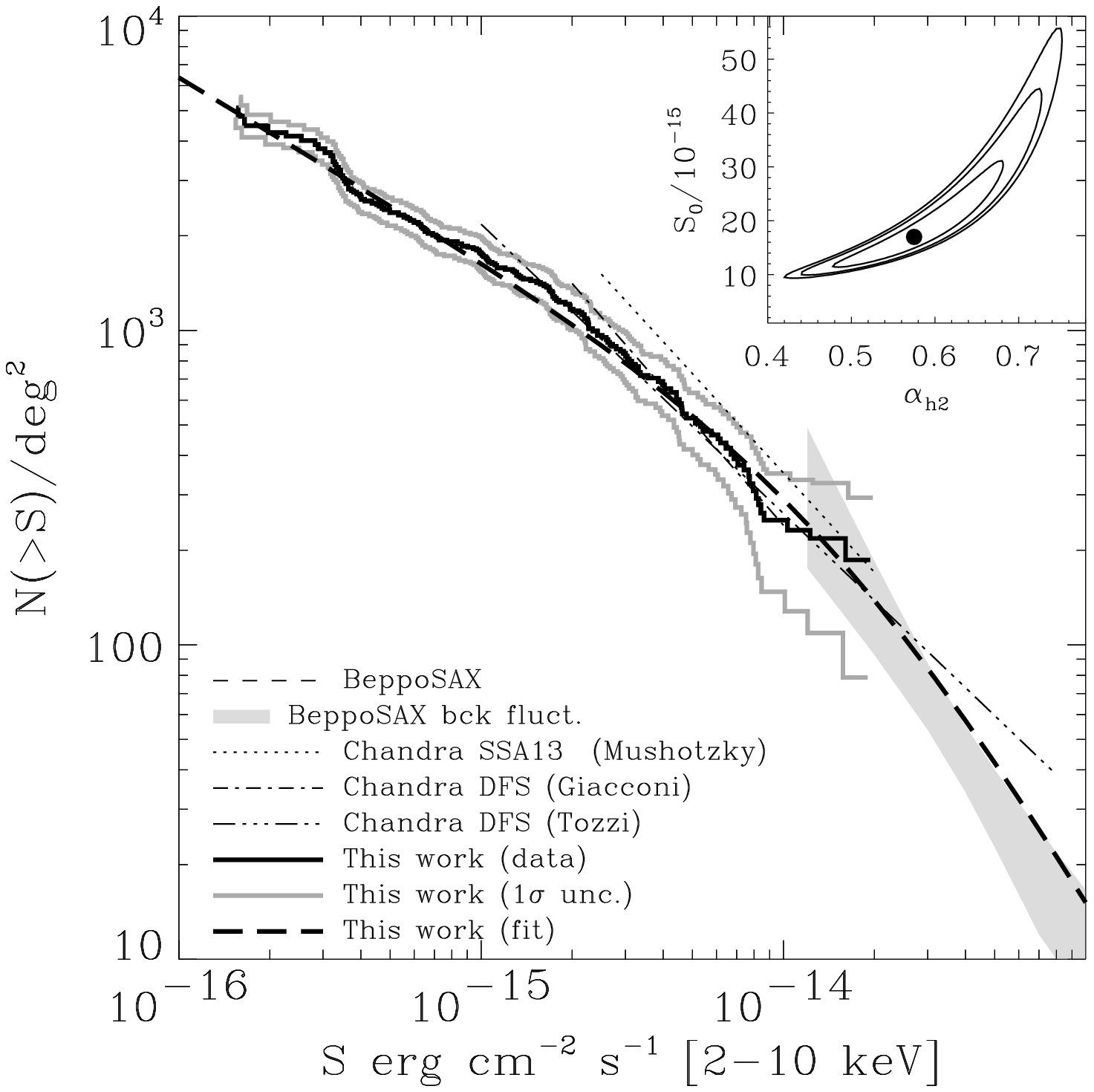}} 
\caption{
{Hard band Log N--Log S from inner $8'$ radius of the 1~Ms observation of 
the CDFS (thick solid line) in the range
$2\times10^{-16}-2\times10^{-14}$~erg~cm$^{-2}$~s$^{-1}$ for an
average $\Gamma=1.4$ power--law spectrum. The distribution is
normalized to the ASCA value and slope (della Ceca et al. 2000). The
gray thick line shows the $1\sigma$ confidence region taking into
account the number statistics and the flux conversion
uncertainties. The thick dashed solid line shows the best M--L fit
smoothly joined broken power law, while the other lines shows fits
from previous works.  The thin dashed line shows the {\it Beppo}SAX
result (Vecchi et al. 1999) while the shaded area shows the result of
the background fluctuation analysis of {\it Beppo}SAX data (Perri \&
Giommi 2000). Lines related to Chandra fields are as in
Fig. \ref{fig:lnlss}. The inset shows the 1, 2 and $3\,\sigma$
contours from the simultaneous M--L fit of the faint slope
$\alpha_{h2}$ and break flux $S_0$, holding the bright slope fixed to
the ASCA result and the normalization fixed to the best--fit value
(see text).}
\label{fig:lnlsh}}
\end{inlinefigure}

\medskip

\section{Conclusions} 

The origin of the XRB has been extensively studied since its discovery
(Giacconi et al. 1962). In the case of the soft (1--2 keV) XRB there
is general consensus that the bulk of the emission can be ascribed to
broad line AGN (i.e. Seyfert 1 galaxies; Hasinger et al. 1998; Schmidt
et al. 1998). At faint fluxes ($\lsim 10^{-15}$ erg s$^{-1}$
cm$^{-2}$) nearby ($z\lsim 0.6$) optically-normal galaxies are also
being detected as soft sources (Barger et al. 2001; Tozzi et al. 2001;
Brandt et al. 2001; Schreier et al. 2001).  Our analysis of the CDFS
extends these studies down to $2\times 10^{-17}$ erg s$^{-1}$
cm$^{-2}$. We found that even at these low fluxes the Log N--Log S
distribution can still be represented by the extrapolation from higher
fluxes (Tozzi et al. 2001) and we are able to resolve
in point sources $>80\%$ of the soft XRB, the main uncertainties being
the sky flux itself.

In the hard band the XRB is made by the superposition of absorbed and
unabsorbed sources (Setti \& Woltjer 1989). In particular, the flat
spectrum of the background ($\Gamma\sim 1.4$) implies a considerable
absorption in most objects. Several studies have emphasized the
importance of the hard XRB as a probe of the evolution of the black
hole population (Fabian \& Iwasawa 1999). The small number of hard
sources not detected in the soft band indicate that even in highly
obscured AGN a sizeable soft emission can still be produced due to,
e.g., scattering, partial covering of the central radiation or from
starburst emission around the AGN (e.g. Turner et al. 1997). This
confirms that also at these extreme flux levels the ratio between soft
and hard X--ray emission is $\sim 1-10\%$. In addition, as shown in Barger et
al. (2001), the effects of obscuration get significantly reduced at
$z\sim 2$, since we observe in the soft band the harder 1.5--6 keV
rest frame interval.

When matching the Log N--Log S distribution from ASCA/{\it{Beppo}}SAX
with deeper data the presence of a knee is evident. We are able to
constrain the Log N--Log S relation below and above the knee, which
occurs at $\sim 2\times 10^{-14}$ erg s$^{-1}$ cm$^{-2}$. We found a
fit to the data with a smoothly joined power law with the bright end
fixed to the ASCA value (1.67) and the fainter one to $\alpha_{\rm
h2}\sim 0.6$. At a level of $\sim 2\times 10^{-16}$ erg s$^{-1}$
cm$^{-2}$ we are fully consistent with the UHURU/HEAO-1 estimate,
whereas we make up $\sim 68\%$ of the hard XRB, if we assume the {\it
Beppo}SAX/ASCA value.

A dedicated, multiwavelenght follow up is mandatory to exploit
the enormous potential of this (and similar) datasets (Barger et
al. 2001; Giacconi et al. 2001b).  The identification of X--ray
sources at the faintest limit and their optical characterization (by
means of photometric redshift) can solve the issue of the
evolution of obscured AGN population at very high redshift.


\begin{acknowledgements}
We thank P. Giommi, G. Ghisellini and E. Bertone for useful comments
and discussions. We thank the referee for suggestions which
improved the paper. We thank the continuous support of the Chandra
Help Desk and the CIAO team for the organization of Chandra/CIAO
workshop. This work was supported through CNAA, Co-fin and ASI
grants.
\end{acknowledgements}

\end{document}